\documentclass[letterpaper, 10 pt, conference]{ieeeconf}
\IEEEoverridecommandlockouts                              
\usepackage{cite}
\usepackage{amsmath,amssymb,amsfonts}
\usepackage{algorithmic}
\usepackage{graphicx}
\usepackage{textcomp}
\usepackage{xcolor}
\usepackage{bm}
\usepackage[capitalise]{cleveref}
\usepackage{subfiles}
\usepackage{array, threeparttable, booktabs} 
\newcommand{\ra}[1]{\renewcommand{\arraystretch}{#1}}
\usepackage{etoolbox}
\usepackage{cite}
\usepackage{multirow}
\makeatletter
\patchcmd{\@makecaption}
  {\scshape}
  {}
  {}
  {}
\makeatother

\def\biblio{\bibliographystyle{IEEEtran}\bibliography{refs}}

\title{\LARGE \bf 
Improved Battery State Estimation Under Parameter Uncertainty Caused by Aging Using Expansion Measurements}

\author{Sravan Pannala, Puneet Valecha, Peyman Mohtat, Jason~B.~Siegel, and Anna~G.~Stefanopoulou%
 \thanks{S. Pannala, P. Mohtat, J.B. Siegel, \& A.G. Stefanopoulou are with the Department of Mechanical Engineering, University of Michigan, Ann Arbor, MI 48109
{\scriptsize \texttt{\{spannala,pmohtat,siegeljb,annastef\}@umich.edu}}}%
 \thanks{P. Valecha is with the Powertrain OBD Calibration, Fiat Chrysler Automotive, Chelsea, MI
{\scriptsize \texttt{valechap@umich.edu}}}
}

\begin{document}
\def\biblio{}

\maketitle
\thispagestyle{empty}
\pagestyle{empty}

\begin{abstract}
Accurate tracking of the internal electrochemical states of lithium-ion battery during cycling enables advanced battery management systems to operate the battery safely and maintain high performance while minimizing battery degradation. To this end, techniques based on voltage measurement have shown promise for estimating the lithium surface concentration of active material particles, which is an important state for avoiding aging mechanisms such as lithium plating. However, methods relying on voltage often lead to large estimation errors when the model parameters change during aging. In this paper, we utilize the in-situ measurement of the battery expansion to augment the voltage and develop an observer to estimate the lithium surface concentration distribution in each electrode particle. We demonstrate that the addition of the expansion signal enables us to correct the negative electrode concentration states in addition to the positive electrode. As a result, compared to a voltage only observer, the proposed observer can successfully recover the surface concentration when the electrodes' stoichiometric window changes, which is a common occurrence under aging by loss of lithium inventory. With a 5\% shift in the electrodes' stoichiometric window, the results indicate a reduction in state estimation error for the negative electrode surface concentration. Under this simulated aged condition, the voltage based observer had 9.3\% error as compared to the proposed voltage and expansion observer which had 0.1\% error in negative electrode surface concentration.
\end{abstract}

\section{Introduction}

Lithium-ion batteries are ubiquitous in our portable computing devices and are playing a major role in the future of transportation with the transition to electric vehicles. To maintain a balance between power/energy demands and cost it is important to have an advanced battery management system that operates the battery safely, close to its limits, while minimizing the degradation. Accurate models and state estimation techniques are required to achieve this performance. The battery models can be classified as Equivalent circuit models (ECMs) and Electrochemical models (EMs). ECMs are widely used in battery management system of electric vehicles because of their computational efficiency and state estimation using ECMs has been widely investigated \cite{plett2004extended,gao2011battery}.

Electrochemical models describe the chemical phenomena occurring inside the battery and thus capture the internal states of a battery, making them suitable for advanced battery control algorithms. Constraints on internal states like negative solid-surface concentration are required to prevent degradation mechanisms like Lithium plating during high C-rate charging.  Full order EMs like Doyle-Fuller-Newman model predict these internal states accurately, but at the expense of computational effort. Reduced order models, like Single Particle Model (SPM), are often used in battery state estimation and control. The SPM assumes a uniform current density across the electrodes and neglecting the electrolyte dynamics, and thus the electrode can be modeled as a single representative spherical particle. More recently, the SPM with electrolyte dynamics (SPMe) has been developed which gives better prediction accuracy compared to SPM.

 Various observers haven been developed for these reduced order models \cite{stetzel2015electrochemical,tang2017state}. In these observers voltage measurement is used to estimate the positive electrode states and the negative electrode states are indirectly calculated by using conservation of lithium in the battery \cite{moura2017battery}, since the positive electrode states are more observable from the voltage measurement \cite{smith2009model}. Thus, these observers are prone to large estimation errors when there is model parameter drift due to aging as the assumption of conservation of lithium no longer holds and hence a single measurement of voltage is insufficient to determine both electrode states \cite{di2010lithium}.

Lithium intercalation and de-intercalation results in volumetric changes in both electrodes of a Li-ion battery and depends on the concentration distribution across the electrodes. These changes can be measured either by the bulk force \cite{mohan2014phenomenological} or expansion measurements \cite{mohtat2020differential} and provide better means to estimate the State of Charge \cite{figueroa2020leveraging} and the State of Health \cite{samad2016battery} of the battery. There are challenges in utilizing mechanical measurement which include difficulties in instrumenting the force/expansion sensors in packs and additional sensor cost.   

This paper develops an observer which uses voltage, expansion and temperature measurements to estimate the individual electrode particle concentrations. We build on the state estimators based on voltage error injection to estimate the concentration in positive electrode particle proposed in  \cite{moura2017battery}, and augment the algorithm using the expansion error injection to estimate the negative electrode particle concentration. With the increase in the number of measurement signals, improvement of the estimator's performance under certain types of model parameter changes was achieved.

\biblio

\section{Model Development}

The battery model presented in this paper is based on the SPM with electrolyte. Additionally a lumped thermal model and concentration dependent expansion is considered. 
\subsection{Single Particle Model with Electrolyte} \label{subsec:SPM}
The SPMe is a commonly used control-oriented electrochemical model for the lithium ion battery. It approximates the full order Doyle-Fuller-Newman (DFN) model under low current operation, where the electrode intercalation reaction is uniform across the electrode thickness and decoupled from changes in electrolyte concentration. In this case, the voltage dynamics are dominated by the solid-phase diffusion of lithium. This solid phase diffusion is modeled by using electrodes with a single representative spherical particle. \cref{eq:diffs_1,eq:diffs_2,eq:diffs_3} show the diffusion equation for a spherical particle along with the requisite boundary conditions at the center and the surface of the particle.
\begin{flalign}
\label{eq:diffs_1}
	\frac{\partial c_{s}}{\partial t}\left(r,t\right)=\frac{1}{r^2}\frac{1}{\partial r}\left[D_{s}r^2\frac{\partial c_{s}}{\partial r}\left(r,t\right)\right] 
\\ \label{eq:diffs_2}
	\frac{\partial c_{s}}{\partial r}\left(0,t\right)=0
\\ \label{eq:diffs_3}
	D_{s}\frac{\partial c_{s}}{\partial r}\left(R_{p},t\right)=-j\left(t\right)\\
	\label{eq:diffs_j}
    j=\frac{I}{a_slA}
\end{flalign}
Here $j$ is the intercalation current density which is given by \cref{eq:diffs_j}, where $A$ is the area, $l$ is thickness of the electrode and $a_s=3\varepsilon_s/R_p$ is the surface area to volume ratio of active material particles. We then use the Bulter-Volmer equation \cref{eq:bv_1} to solve for the overpotential of the intercalation reaction $\eta$, where $i_{0}$ is the exchange current density \cref{eq:bv_2}, $c_{ss}(t)=c_{s}(R_{p},t)$ is the concentration at the surface of the particle, the $k_0$ is the reaction rate constant, and the ($\alpha_a, \alpha_c$) are the charge transfer coefficients. 
\begin{flalign}
\label{eq:bv_1}
	j\left(t\right)=\frac{i_{0}\left(t\right)}{F}\left(e^{\frac{\alpha_aF}{RT}\eta}-e^{\frac{-\alpha_cF}{RT}\eta}\right)
\\ \label{eq:bv_2}
	i_{0}\left(t\right)=k_0\left(\bar{c_e}\left(t\right)\right)^\alpha\left(c_{s,max}-c_{ss}\left(t\right)\right)^\alpha\left(c_{ss}\left(t\right)\right)^\alpha
\end{flalign}

The electrolyte diffusion equations are derived based on the assumptions in \cite{moura2017battery} with boundary conditions: the continuity of $c_{e}$, and $\nabla c_e(0,t)=\nabla c_e(l^t,t)=0$.
\begin{multline}
\label{eq:diffe}
\epsilon_e \frac{\partial c_{e}}{\partial t}(x,t)=\nabla.(D_e^{eff} \nabla c_e(x,t))\\+\frac{1-t_+^0}{F}\times
\begin{cases}
\frac{I(t)}{l^-} & 0\le x<l^-,\\
0 & l^-\le x\le l^-+l^s ,\\
\frac{-I(t)}{l^+} & l^-+l^s<x\le l^t,\\
\end{cases}
\end{multline}
where $l^t=l^-+l^s+l^+$. The liquid-phase Ohm's law is shown in \cref{eq:ohml_1}.
\begin{multline}
\label{eq:ohml_1}
i_e(x,t) = -\kappa^{eff}\nabla \Phi_e(x,t) + \frac{2\kappa^{eff}RT}{F}(1-t_+^0)\\
\left( 1+\frac{d~ln~f_{\pm}}{d~ln~c_e}(x,t)  \right)\nabla (ln~c_e) (x,t)
\end{multline}
Integrating and applying the boundary condition results in
\begin{multline}
\Phi_e(l,t)= - \left( \frac{l^-}{2(\epsilon^-)^{brugg}}+ \frac{l^s}{(\epsilon^s)^{brugg}} +\frac{l^+}{2(\epsilon^+)^{brugg}}\right)\\\frac{I(t)}{\kappa} +\frac{2RT}{F}(1-t_+^0)t_f(ln ~ c_e(l^t,t)-ln ~ c_e(0,t))
\label{eq:ohml_2}
\end{multline}
where the concentration dependence of the $\kappa$ is neglected for simplicity, and the term $t_f=\left( 1+\frac{d~ln~f_{\pm}}{d~ln~c_e}(x,t)  \right)$ is assumed to be constant. 

The initial concentrations of the electrodes are given by
\begin{flalign}
\label{eq:csp_init}
c_{s,0}^+=c_{s,max}^+(SOC_0\times(y_{100}-y_0)+y_0)\\
\label{eq:csn_init}
c_{s,0}^-=c_{s,max}^-(SOC_0\times(x_{100}-x_0)+x_0)
\end{flalign}
where $SOC_0$ is the initial state of charge, $c_{s,max}$ is the maximum particle concentration, $y_{100},\,y_0$ are the positive electrode stoichiometric windows and $x_{100},\,x_0$ are the negative electrode stoichiometric windows defined by the voltage limits and electrode physical dimensions \cite{mohtat2019towards}.

Finally the terminal voltage of the battery is given by \cref{eq:vt} where $U$ is the half-cell open circuit potential, and $V_{R}(x,t)=R_{f}Fj(x,t)$ is the voltage drop due to the film resistance.
\begin{multline}
\label{eq:vt} 
V_{t}(t)= h_v(c_{ss}^+,c_{ss}^-,c_e,I(t))=\eta^+(t)+U^+(c_{ss}^+(t))\\+V_{R}^+(t)-\eta^+(t)\-U^-(c_{ss}^-(t))-V_{R}^-(t)+\Phi_e(l^t,t)
\end{multline}

\subsection{Thermal Model} \label{subsec:thmodel}

The thermal model used in this paper is a one-state lumped model for battery temperature,  
\begin{multline}
\label{eq:th}
C_{th} \frac{dT_b}{dt}(t) =-h(T_b(t)-T_{a}(t))\\+I(t)(U^+(c_{ss}^+(t))-U^-(c_{ss}^-(t))-V_t(t))
\end{multline}
where $C_{th}$ is the lumped heat capacity, $T_a$ is the ambient air temperature, $h$ is the heat transfer coefficient, and the only source of heat generation inside the battery is joule heating. The effect of Entropic heating can be ignored at the C-rates of interest.

\subsection{Expansion Model} \label{subsec:exp}
The expansion model used in the paper is based on the model used in \cite{mohtat2020differential}. 
\subsubsection{Intercalation induced expansion} \label{subsec:intcl_exp}
The displacement at the surface of the particle is obtained by solving stress strain relationship in the particle with intercalation expansion as detailed in \cite{mohtat2020differential} is given by \cref{eq:ur}.
\begin{flalign}
\label{eq:ur}
u_R(t)=\frac{1}{(R_{p})^2}\int_0^{R_{p}} \rho ^2 \Delta \mathcal{V}\left(c_s(\rho,t)\right)d\rho
\end{flalign}
where $\Delta\mathcal{V}\left(c_s(r)\right)$ is the particle expansion function in terms of volumetric strain.
\subsubsection{Electrode Expansion} \label{subsec:el_exp}
The electrode in a Li-battery is made of active material, binder and conductive material. In our model we assume that the expansion of electrode components other than the active material to be negligible. We further assume that the electrode only expands in the through-plane direction. Using the displacement at the surface of particle shown in \cref{eq:ur} and the above assumptions, we obtain the change in electrode thickness:
\begin{flalign}
\label{eq:dt_el}
\Delta t=a_s l u_R(t) 
\end{flalign}
\subsubsection{Thermal Expansion} \label{subsec:th_exp}
The lumped thermal model in \cref{subsec:thmodel}, is used to predict the thermal expansion, which is given by
\begin{flalign}
\label{eq:dt_th}
\Delta t_{th}=\alpha_{th}\left(T_b-T_0\right).
\end{flalign}
Here $\alpha_{th}$ is the thermal expansion coefficient and $T_0$ is the reference temperature, and $T_b$ is the battery temperature given by Eq.~\ref{eq:th}.

\subsubsection{Total Expansion} \label{subsec:tot_exp}
The total electrode expansion is the sum of the expansion of individual electrodes. Pouch cell Li-ion batteries contain multiple layers, So the single layer expansion is multiplied by the number of layers to find the total expansion. Also the cell level expansion is influenced by separator, current collectors and casing. The elasticity of these layers are approximated with a linear spring. The total electrode expansion is given by \cref{eq:dt_tel}, where $\kappa_b$ is tuning parameter.
\begin{flalign}
\label{eq:dt_tel}
\Delta t_{e}=\kappa_b(\Delta t^++\Delta t^-)
\end{flalign}
Now we calculate total battery expansion:
\begin{flalign}
\label{eq:dt_t}
\Delta t_{b}=\Delta t_e+\Delta t_{th}
\end{flalign}
where the total expansion is calculated by adding the total electrode expansion and the thermal expansion.
\biblio

\section{Observer Design}

\begin{figure} [!h]
	\centering
	\includegraphics[width=0.9\linewidth]{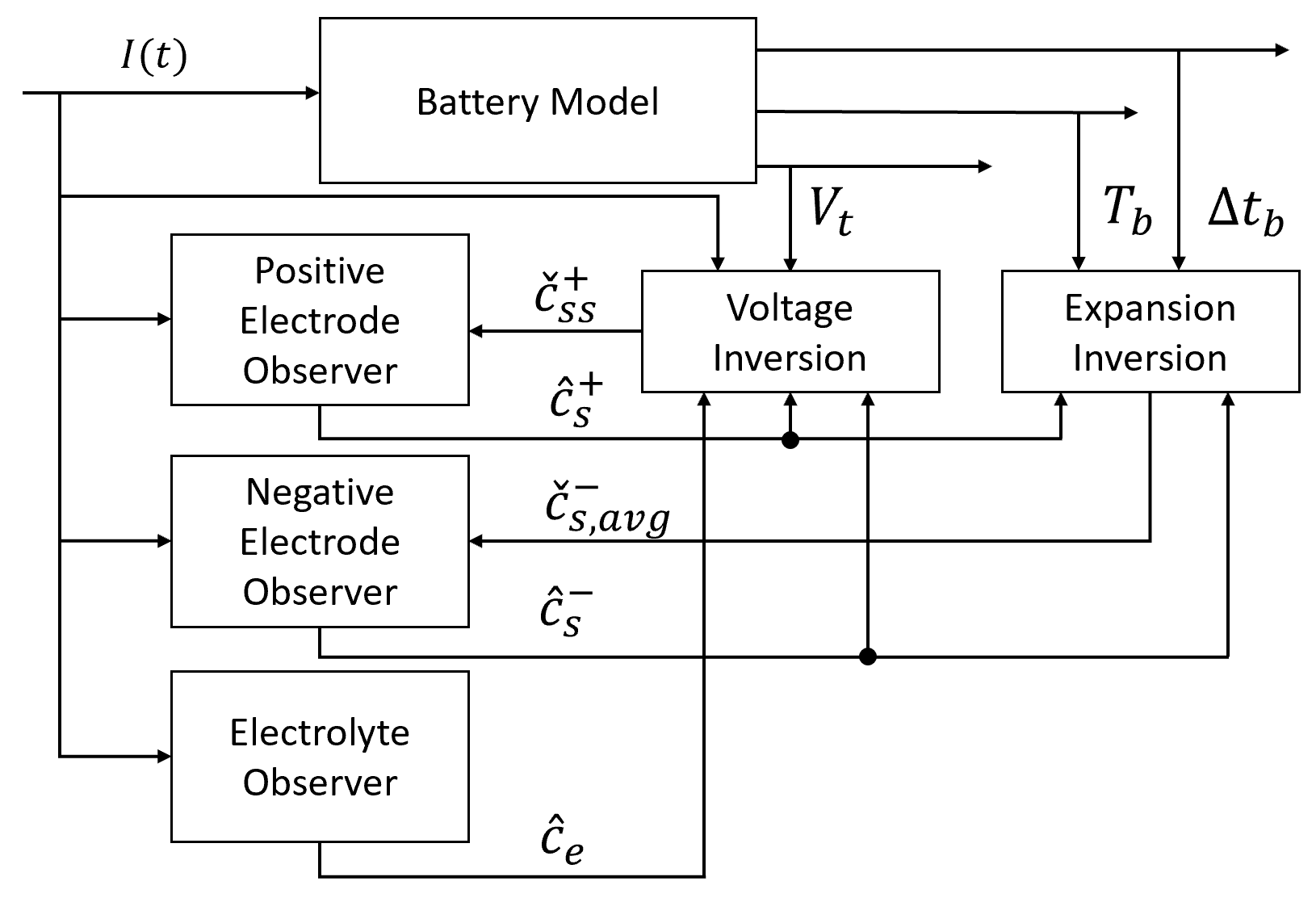}
	\caption{Observer Schematic Diagram. The positive electrode observer depends on the inverted surface concentration \(\check{c}_{ss}^+\) from the voltage inversion block which uses measured battery terminal voltage \(V_t\). The voltage inversion depends on the open loop Electrolyte concentration estimate \(\hat{c}_e\). The estimated positive electrode concentration \(\hat{c}_s^+\) is then used in the expansion inversion block in combination with the measured battery temperature \(T_b\) and expansion \(\Delta t_b\) to inform the negative electrode observer \(\hat{c}_s^-\) using inverted negative electrode average concentration \(\check{c}_{s,avg}^-\).}
	\label{fig:obs_sch}
\end{figure}

The block diagram of the observer is shown in \cref{fig:obs_sch}. \cref{subsec:cathobs,subsec:volinv,subsec:elobs} are adopted from \cite{moura2017battery} and are briefly described below.

\subsection{Positive Electrode Observer} \label{subsec:cathobs}
The positive electrode observer uses a copy of model and injects boundary state error as shown in \cref{eq:caobs_1,eq:caobs_2,eq:caobs_3}
\begin{multline}
\label{eq:caobs_1}
\frac{\partial{\hat{c}}_s^+}{\partial t}\left(r,t\right)=D_s^+\left[\frac{2}{r}\frac{\partial{\hat{c}}_s^+}{\partial r}\left(r,t\right)+\frac{\partial^2{\hat{c}}_s^+}{\partial r^2}\left(r,t\right)\right]\\+\overline{p}^+(r)\ [{\check{c}}_{ss}^+-{\hat{c}}_{ss}^+]
\end{multline}
\begin{flalign}
\label{eq:caobs_2}
\frac{\partial{\hat{c}}_s^+}{\partial r}\left(0,t\right)=0
\end{flalign}
\begin{flalign}
\label{eq:caobs_3}
\frac{\partial{\hat{c}}_s^+}{\partial r}\left(R_p^+,t\right)=\frac{I(t)}{D_s^-Fa_s^-l^-}+\overline{p}_0^+\ [{\check{c}}_{ss}^+-{\hat{c}}_{ss}^+]
\end{flalign}
where $\check{c}_{ss}^+$ is inverted surface concentration calculated using \cref{eq:vtinv_4}. The observer gains are derived with the backstepping approach:
\begin{flalign}
\label{eq:obsgain_1}
\overline{p}^+(r)=\frac{-\lambda D_s^+}{2R_p^+\overline{z}}\left[I_1(\overline{z})-\frac{2\lambda}{\overline{z}}I_2(\overline{z})\right]
\end{flalign}
\begin{flalign}
\label{eq:obsgain_2}
\overline{z}=\sqrt{\lambda\left(\frac{r^2}{(R_p^+)^2}-1\right)}
\end{flalign}
\begin{flalign}
\label{eq:obsgain_3}
\overline{p}_0^+=\frac{1}{2R_p^+}(3-\lambda), \quad \textrm{for} \: \lambda<\frac{1}{4}
\end{flalign}
where $I_1(z)$ and $I_2(z)$ are first and second order modified Bessel functions of the first kind, and $\lambda$ controls the eigenvalue locations and determines the convergence rate. 

\subsection{Voltage Inversion} \label{subsec:volinv}
In this section we use a nonlinear gradient algorithm which estimates $\check{c}_{ss}^+$ by inverting the nonlinear $V_t$ output function given in \cref{eq:vt}. 
\begin{flalign}
\label{eq:vtinv_1}
V_t(t)=h_v(c_{ss}^+,t)
\end{flalign}
The dependency of this nonlinear output function on $c_{ss}^-$, $\Phi_e$ and $I(t)$ is suppressed to a single dependence on $t$.
We now define inversion error signal $e_{V}(t)$ in \cref{eq:vtinv_2} and regressor signal $\phi_v(t)$ in \cref{eq:vtinv_3}.
\begin{flalign}
\label{eq:vtinv_2}
e_v(t)=V_t(t)-h_v(\check{c}_{ss}^+,t)\\
\label{eq:vtinv_3}
\phi_v(t)=\frac{\partial h_v}{\partial c_{ss}^+}(\check{c}_{ss}^+,t)
\end{flalign}
Gradient update law for $\check{c}_{ss}^+$ is given by \cref{eq:vtinv_4}, where $\gamma_v$ is a tuning parameter.
\begin{flalign}
\label{eq:vtinv_4}
\frac{d}{dt}\check{c}_{ss}^+=\gamma_v\phi_v(t)e_v(t)
\end{flalign}

\subsection{Electrolyte Observer} \label{subsec:elobs}
The electrolyte observer used is a open-loop observer which has the same form as the model. The equations of the observer are provided in \cref{eq:obs_diffe} with boundary conditions: the continuity of $\hat{c}_{e}$, and $\nabla \hat{c}_e(0,t)=\nabla \hat{c}_e(l,t)=0$.
\begin{multline}
\label{eq:obs_diffe}
\epsilon_e \frac{\partial \hat{c}_{e}}{\partial t}(x,t)=\nabla.(D_e^{eff} \nabla \hat{c}_e(x,t))\\+\frac{1-t_+^0}{F}\times
\begin{cases}
\frac{I(t)}{l^-} & 0\le x<l^-,\\
0 & l^-\le x\le l^-+l^s ,\\
\frac{-I(t)}{l^+} & l^-+l^s<x\le l,\\
\end{cases}
\end{multline}

\subsection{Negative Electrode Observer} \label{subsec:anobs}
The negative electrode observer uses a copy of model and injects $c_{s,avg}^-$ error as shown in \cref{eq:anobs_1,eq:anobs_2,eq:anobs_3}
\begin{multline}
\label{eq:anobs_1}
\frac{\partial \hat{c}_s^-}{\partial t}\left(r,t\right)=D_s^-\left[\frac{2}{r}\frac{\partial{\hat{c}}_s^-}{\partial r}\left(r,t\right)+\frac{\partial^2{\hat{c}}_s^-}{\partial r^2}\left(r,t\right)\right]\\+k^-\left[{\check{c}}_{s,avg}^--{\hat{c}}_{s,avg}^-\right]
\end{multline}
\begin{flalign}
\label{eq:anobs_2}
\frac{\partial{\hat{c}}_s^-}{\partial r}\left(0,t\right)=0
\end{flalign}
\begin{flalign}
\label{eq:anobs_3}
\frac{\partial{\hat{c}}_s^-}{\partial r}\left(R_p^-,t\right)=\frac{-I(t)}{D_s^-Fa_s^-l^-}
\end{flalign}
where $k^-$ is the feedback gain which determines the system stability and convergence rate. Note by comparison of \cref{eq:anobs_3,eq:caobs_3}, the anode observer does not adjust the estimate of the concentration gradient, only the average value, and relies on the open loop dynamics for prediction of the concentration gradient.

\subsection{Expansion Inversion} \label{subsec:expinv}
We use the expansion measurement $\Delta t_b$ and the temperature measurement $T_b$ to estimate the average negative electrode concentration $\check{c}_{s,avg}^-$. The steps followed are described below.
\subsubsection{Estimating negative electrode particle displacement} \label{subsec:expinv_ur}
We start by first estimating the thermal expansion by using the battery temperature measurement.    
\begin{flalign}
\label{eq:expinv_th}
\Delta \hat{t}_{th}=\alpha_{th}\left(T_b-T_0\right)
\end{flalign}
 Then we use the positive electrode observer states $\hat{c}_s^+(t)$ to estimate positive electrode expansion $\Delta t^+$.
\begin{flalign}
\label{eq:expinv_ca_1}
\hat{u}_R^+(t)=\frac{1}{(R_{p}^+)^2}\int_0^{R_{p}^+} \rho ^2 \Delta \mathcal{V}\left(\hat{c}_s^+(\rho,t)\right)d\rho\\
\label{eq:expinv_ca_2}
\Delta \hat{t}^+=a_s^+ l^+ \hat{u}_R^+(t)
\end{flalign}
Both of these estimates are used to estimate the negative electrode expansion as shown in \cref{eq:expinv_an}. 
\begin{flalign}
\label{eq:expinv_el}
\Delta \check{t}_e=\kappa_b\left(\Delta \check{t}^-+\Delta \hat{t}^+\right)=\Delta t_{b}-\Delta \hat{t}_{th}\\
\label{eq:expinv_an}
\Delta \check{t}^-=\frac{\Delta t_{b}-\Delta \hat{t}_{th}}{\kappa_b}-\Delta \hat{t}^+
\end{flalign}
Finally the particle displacement at the surface is by 
\begin{flalign}
\label{eq:ur_inv}
\check{u}_R^-(t)=\frac{\Delta \check{t}^-}{a_s^-l^-}.
\end{flalign}




\subsubsection{Estimating negative electrode average concentration}
In this section we develop a way to estimate average negative electrode concentration from negative electrode particle displacement. To start, we first define a new variable $\Tilde{c}_s^-$ in \cref{eq:cst}.
\begin{flalign}
\label{eq:cst}
\Tilde{c}_s^-(r,t)=\hat{c}_s^-(r,t)-\hat{c}_{s,avg}^-
\end{flalign}
where $\hat{c}_{s,avg}^-$ is the average negative electrode concentration of the observer states calculated in \cref{subsec:anobs} . We now use the $\Tilde{c}_s^-(r,t)$, $\check{u}_R^-(t)$ from \cref{eq:ur_inv} and \cref{eq:ur} to estimate the inverted negative electrode average concentration $\check{c}_{s,avg}^-(t)$, by solving
\begin{multline}
\label{eq:nlexp_inv}
\check{u}_R^-(t) = h_e(\Tilde{c}_s^-(r,t)+{c}_{s,avg}^-(t))\\
=\frac{1}{(R_{p}^-)^2}\int_0^{R_{p}^-} \rho ^2 \Delta \mathcal{V}\left(\Tilde{c}_s^-(\rho,t)+{c}_{s,avg}^-(t)\right)d\rho.
\end{multline}

To solve \cref{eq:nlexp_inv} we implement a gradient update law similar to voltage inversion in \cref{subsec:volinv}. We now define inversion error signal $e_{e}(t)$ in \cref{eq:nlexp_1} and regressor signal $\phi_e(t)$ in \cref{eq:nlexp_2}. Gradient update law for $\check{c}_{s,avg}^-$ is given by \cref{eq:nlexp_3}, where $\gamma_e$ is a tuning parameter.
\begin{flalign}
\label{eq:nlexp_1}
e_e(t)=\check{u}_R^-(t)-h_e(\Tilde{c}_s^-(r,t)+\check{c}_{s,avg}^-(t))\\
\label{eq:nlexp_2}
\phi_e(t)=\frac{\partial h_e}{\partial {c}_{s,avg}^-}(\Tilde{c}_s^-(r,t)+\check{c}_{s,avg}^-(t))\\
\label{eq:nlexp_3}
\frac{d}{dt}\check{c}_{savg}^-=\gamma_e\phi_e(t)e_e(t)
\end{flalign}
This introduces a dynamic coupling between the concentration state observers for the positive and negative electrodes.

\biblio

\section{Results and Discussion}

\begin{figure} [h]
	\centering
    \includegraphics[width=\columnwidth]{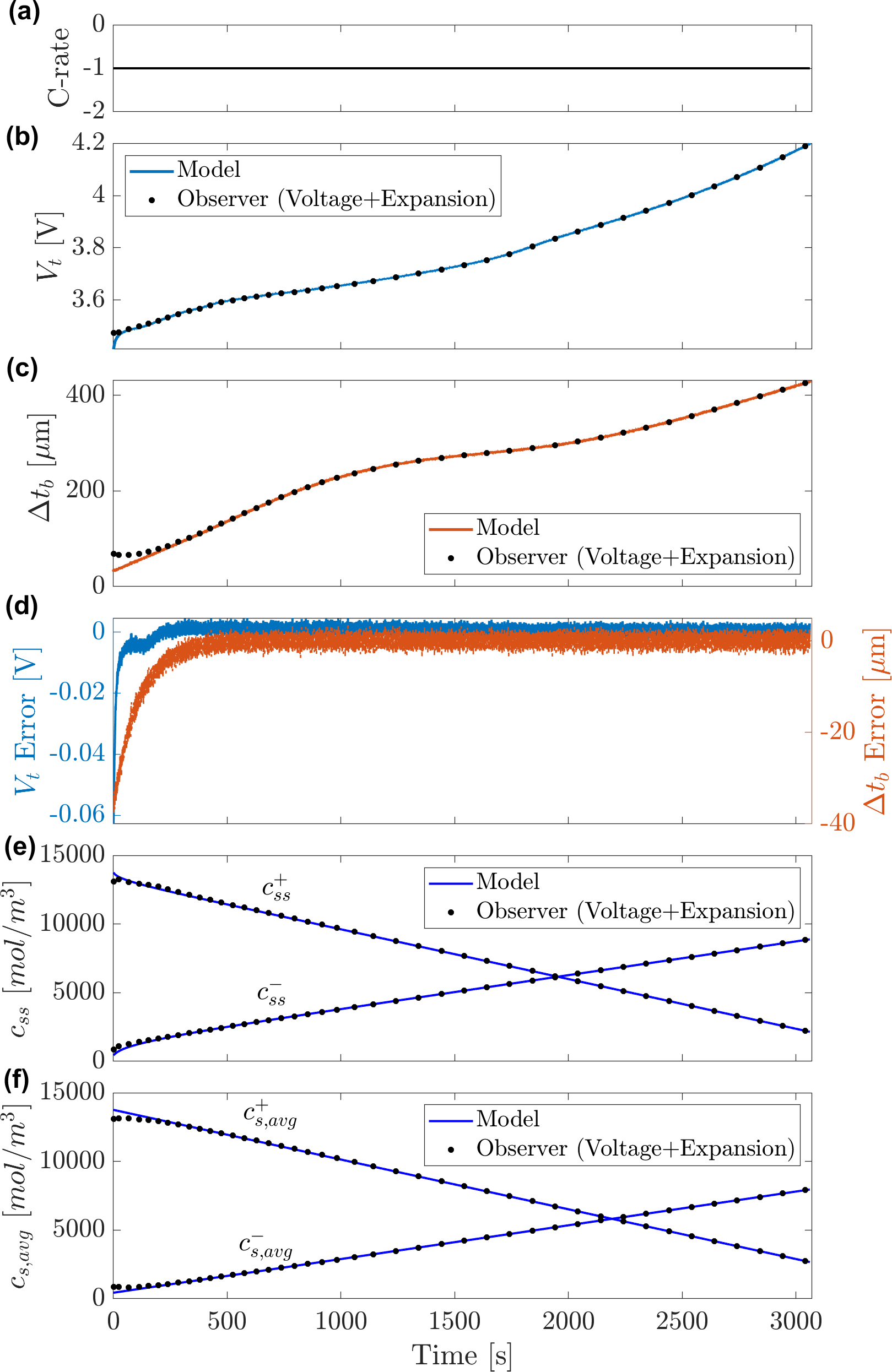}
	\caption{Simulation results for 1C Constant Current input with $SOC_0=0.05$ and $ \hat{SOC}_0=0.10$. (a) Current (b) Voltage (c) Expansion (d) Voltage and Expansion Errors (e) Surface Concentration of both electrodes \(c_{ss}^-,c_{ss}^+\) (f) Average Concentration of both electrodes \(c_{s,avg}^-,c_{s,avg}^+\). Voltage and expansion converge to the measured values within 5 minutes.}
	\label{fig:sim1}
\end{figure}

In this section we present the simulation results of the observer on the plant model. The diffusion equations in the model and observer are discretized using Method of Lines with second-order approximation of the boundary conditions \cite{Versypt2014}. The following observer parameters are used for all simulations; $\gamma_v = 10^8$, $\lambda =-20$, $\gamma_e=10^{22}$ and $k^-=0.01$. Additional noise is added to voltage and expansion signals with a standard deviation of $1\,mV$ for voltage and $1\,\mu m$ for expansion.

\subsection{Constant Current Charge} \label{subsec:results_cc}
First we simulate a constant current charge of 1C. The simulated battery is initialized with $SOC_0=0.05$ and the observer with $ \hat{SOC}_0=0.1$. We can see from \cref{fig:sim1} that $\hat{V}_t$, $\Delta\hat{t}_b$, $\hat{c}_{ss}^-$, $\hat{c}_{s,avg}^-$, $\hat{c}_{ss}^+$ and $\hat{c}_{s,avg}^+$ converge. The terminal voltage $\hat{V}_t$ converges faster than $\Delta\hat{t}_b$ as $\hat{V}_t$ depends on $\hat{c}_{ss}$ convergence but $\Delta\hat{t}_b$ depends on $\hat{c}_{s,avg}$ convergence which is slower. This is because $\hat{c}_{s,avg}$ is a linear combination of all $\hat{c}_{s}(r)$ states and convergence of $\hat{c}_{s,avg}$ depends on convergence of all the states including faster and slower states. We compare the performance of the Voltage and Expansion based observer (referred to as V+EXP-obs) which uses voltage, temperature and expansion measurements for with the one in \cite{moura2017battery}, which uses only voltage measurement (referred to as V-obs). The root mean square percent error (RMSPE) of $\Delta\hat{t}_b$, $\hat{c}_{ss}^-$, $\hat{c}_{s,avg}^-$, $\hat{c}_{ss}^+$ and $\hat{c}_{s,avg}^+$ estimates after five minutes of simulation are given in \cref{tabel1}. While the RMSPE of positive electrode concentration estimates have similar values for both V+EXP-obs and V-obs, the RMSPE of negative electrode concentration estimates is slightly higher for V-obs.


\subsection{Model Drift due to Aging}
As the battery ages a number of parameters in the model drift from their initial values. Hence, it is important to evaluate the observer performance with uncertainty in parameters. There are number of aging mechanisms that contribute to parameter mismatch during aging, namely loss of lithium inventory (LLI) and loss of active material (LAM). It is known that these aging mechanisms affect the battery parameters like stoichiometric windows in negative electrode $x_{100}$ and in positive electrode $y_0$, and active material ratio of negative electrode $\varepsilon_s^-$, which change as the battery ages \cite{mohtat2017identifying}. 
\begin{figure} [!t]
	\centering
    \includegraphics[width=\columnwidth]{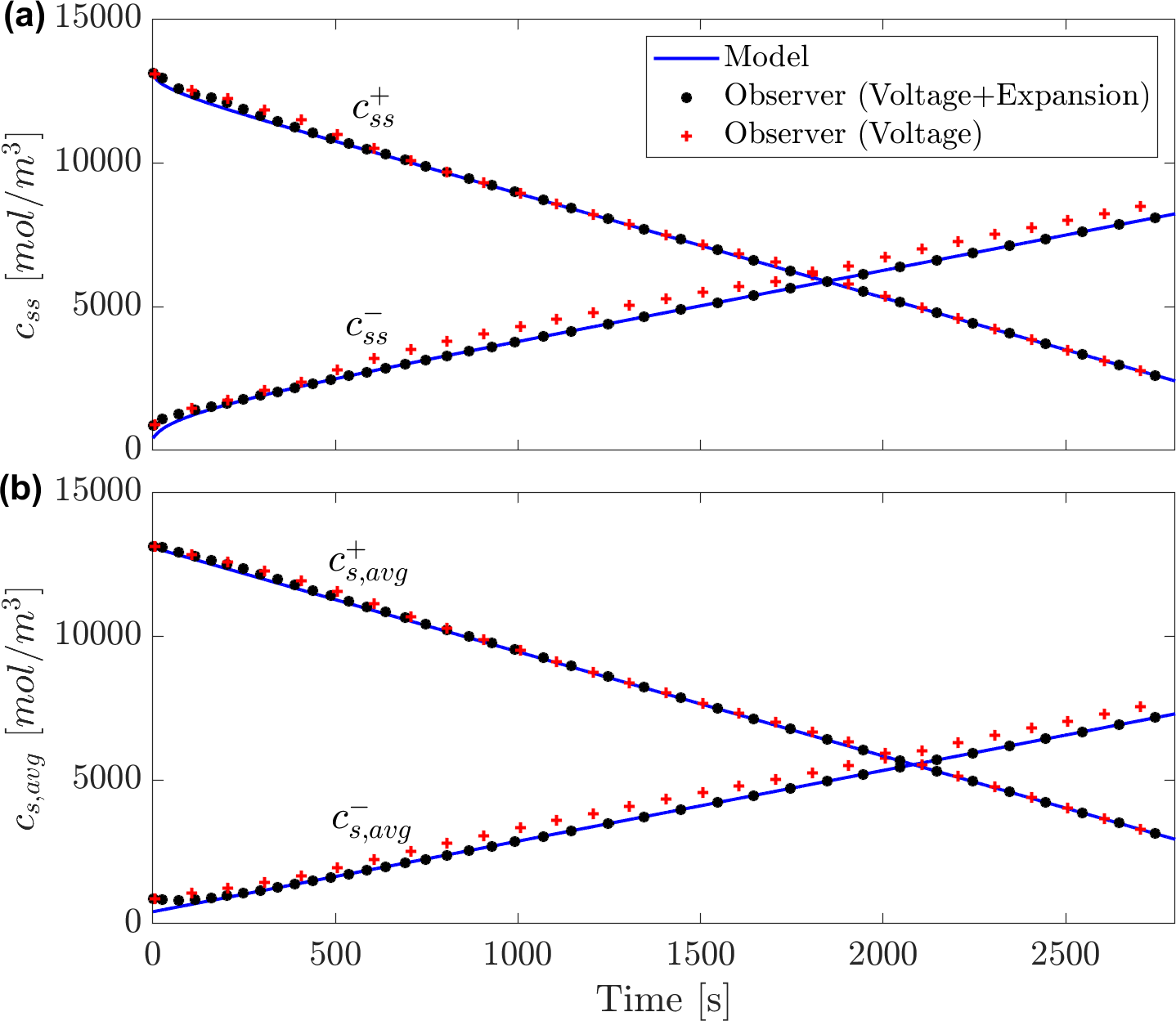}
	\caption{Simulation results for Stoichiometric Window Drift for Voltage and Expansion observer and Voltage only observer during Constant Current charge at 1C rate with $SOC_0=0.05$ and $ \hat{SOC}_0=0.10$. (a) Surface Concentration of both electrodes \(c_{s,s}^-,c_{s,s}^+\) (b) Average Concentration of both electrodes \(c_{s,avg}^-,c_{s,avg}^+\). $\hat{c}_{s}^+$ converges for both, but $\hat{c}_{s}^-$ converges only for V+EXP-obs.}
	\label{fig:sim3}
\end{figure}

\subsubsection{Stoichiometric Window Change}
First we simulate a case where both $x_{100}$ and $y_0$ are reduced by 5\% in the plant due to aging, and the parameters in observer are unchanged. The observer and plant are initialized as in  \cref{subsec:results_cc}. The results of the simulations are shown in \cref{fig:sim3}. We can see that while the $\hat{c}_{ss}^+$ and $\hat{c}_{s,avg}^+$ converge for both observers, $\hat{c}_{ss}^-$ and $\hat{c}_{s,avg}^-$ converges for the V+EXP-obs but not for V-obs. This is because additional feedback in V+EXP-obs compensates for the model mismatch in the negative electrode parameters resulting in better estimation of $\hat{c}_{s}^-$ states, while in V-obs the $\hat{c}_{s}^-$ states are calculated by using Lithium conservation. Also, this higher $\hat{c}_{s}^-$ error in V-obs causes higher error in $\hat{c}_{ss}^+$ as shown in \cref{tabel1}.

\subsubsection{Active Material Loss}
\begin{figure} [!h]
	\centering
    \includegraphics[width=\columnwidth]{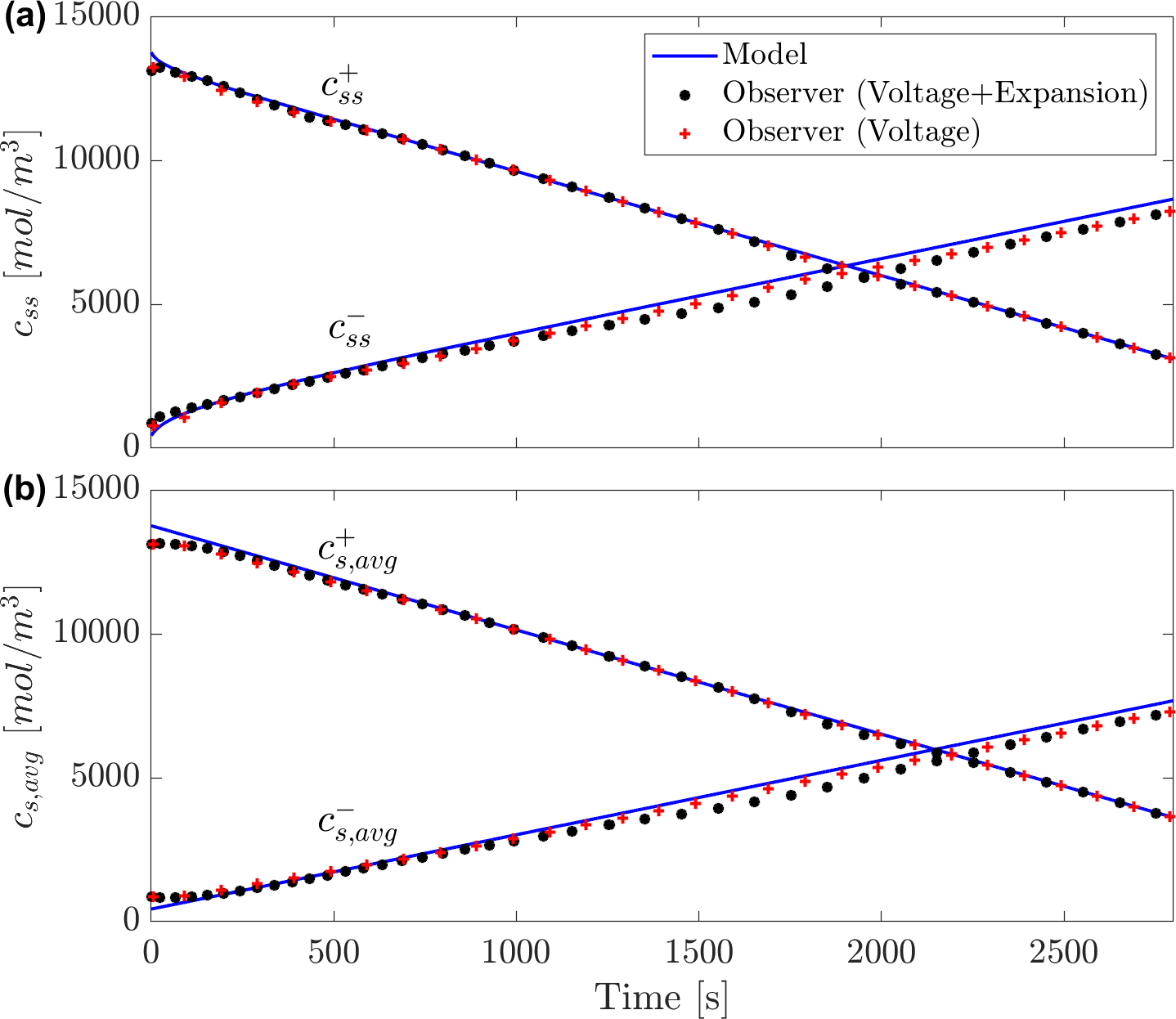}
	\caption{Simulation results for Active Material Ratio Drift for V+EXP-obs and V-obs with $SOC_0=0.05$ and $ \hat{SOC}_0=0.10$. (a) Surface Concentration of both electrodes \(c_{s,s}^-,c_{s,s}^+\) (b) Average Concentration of both electrodes \(c_{s,avg}^-,c_{s,avg}^+\). The estimates of negative electrode concentration states $\hat{c}_{s}^-$ does not converge for either observer.}
	\label{fig:sim4}
\end{figure}

Next we simulate a 5\% parametric error in the negative electrode volume fraction, $\varepsilon_s^-$  due to aging. The observer and simulated battery are initialized as in  \cref{subsec:results_cc}. The results of the simulations are shown in \cref{fig:sim4}. We can see that while the $\hat{c}_{ss}^+$ and $\hat{c}_{s,avg}^+$ converge for both observers, both $\hat{c}_{ss}^-$ and $\hat{c}_{s,avg}^-$ do not converge for V+EXP-obs or V-obs. Also error of $\hat{c}_{s,avg}^-$ is higher in V+EXP-obs compared to V-obs as seen from the RMSPE values given in \cref{tabel1}. This is because $\varepsilon_s^-$ is used in the output function inversion of expansion leading to inaccurate estimation of $\check{c}_{s,avg}^-$, thus resulting in inaccurate estimates of $\hat{c}_{s}^-$ states in V+EXP-obs.

\begin{table}[!h]
\centering
\begin{threeparttable}
\caption{Simulation Error in RMSPE (after 5 minutes) of Concentration Estimates for V+EXP-obs and V-obs}

\label{tabel1}
\ra{0.9}
\begin{tabular}{l c c c c c c}
\toprule
\multicolumn{1}{c}{Estimates} & \multicolumn{6}{c}{Simulation Error (\%)}\\ \cmidrule(lr){1-1} \cmidrule(lr){2-7}
& \multicolumn{2}{c}{\multirow{2}{*}{Fresh Cell}} & \multicolumn{4}{c}{Aged Cell}\\  \cmidrule(lr){4-7}
& & & \multicolumn{2}{c}{Stoich Change \tnote{1}} &  \multicolumn{2}{c}{$\varepsilon_s^-$ Loss \tnote{2}}\\ \cmidrule(lr){2-3} \cmidrule(lr){4-5} \cmidrule(lr){6-7}
& V+E & V & V+E & V & V+E & V \\ \midrule
 $\hat{c}_{ss}^-$ \tnote{\dag} & {0.2} & {1.2} & {0.1} & {9.3} & {6} & {4.6} \\
 $\hat{c}_{s,avg}^-$ \tnote{\S} & {0.4} & {2} & {0.2} & {11.6} & {6.3} & {4} \\
 $\hat{c}_{ss}^+$ \tnote{\dag}& {0.3} & {0.2} & {0.3} & {1.4} & {1.1} & {4.6} \\
 $\hat{c}_{s,avg}^+$ \tnote{\S} & {0.2} & {0.4} & {0.4} & {1.3} & {1} & {4.6} \\
\bottomrule
\end{tabular}
\begin{tablenotes}[flushleft]
      \footnotesize
      \item[1]Stoichiometric window change \item[2] Active material loss
      \item[\dag] Negative/Positive electrode surface concentration 
      \item[\S] Negative/Positive electrode average concentration
    \end{tablenotes}
\end{threeparttable}
\end{table}

\subsection{Summary of Simulation Results}
The outputs concentration state estimation errors for $\hat{c}_{ss}^-$, $\hat{c}_{s,avg}^-$, $\hat{c}_{ss}^+$ and $\hat{c}_{s,avg}^+$, after the initial convergence period, are given in \cref{tabel1}. These errors are calculated with the values after five minutes to simulations to normalize initialization errors across the simulations. The negative electrode concentrations $\hat{c}_{ss}^-$ and  $\hat{c}_{s,avg}^-$ of V+EXP-obs have slightly lower errors for Constant Current simulation compared to V-obs. For the 1C charge simulation for the aged cell with change in Stoichiometric window, the concentration errors for $\hat{c}_{ss}^-$ in V-obs is 9.3\% which is much higher than 0.1\% for V+EXP-obs. Even $\hat{c}_{ss}^+$ is higher in V-obs. For the $\varepsilon_s^-$ loss case all the concentration errors have high values for both the observers. While $\hat{c}_{ss}^-$ of V+EXP-obs has error of 6\% and V-obs has a slightly lower error of 4.6\%, $\hat{c}_{ss}^-$ of V+EXP-obs has a lower error of 1.1\% against 4.6\% of V-obs.

\biblio

\section{Conclusion}

In this paper we have developed a state observer for a physics based single particle Li-ion battery model by augmenting the voltage measurement with expansion measurement. The observer shows improved convergence of the concentration states. The observer performance is also evaluated against parametric modeling error representative of battery aging. This model error causes error in the negative electrode concentration states when using only voltage measurement for state estimation. Although the addition of expansion measurement doesn't improve observer performance in case of negative electrode active material ratio drift, the proposed observer was able to compensate for drift in stoichiometric windows. This can be seen in the error of negative electrode surface concentration which has a high value of 9.3\% for voltage only observer, but has a value of 0.1\% for voltage and expansion observer. Finally, accurate estimation of  negative solid-surface concentration can enable more robust constraints on the state during charging and prevent degradation mechanisms like Lithium plating during high C-rates. 
\addtolength{\textheight}{-12cm} 
\section*{Acknowledgement}

The authors would like to acknowledge the technical and financial support of Mercedes-Benz R\&D North America.

\bibliographystyle{IEEEtran}
\bibliography{refs}

\end{document}